# Magnetoresistance and Hall effect in e-doped superconducting SrLaCuO thin films

V Jovanovic, Z Z Li, F Bouquet, L Fruchter and H Raffy[1]

Laboratoire de physique des Solides, UMR8502-CNRS, Université Paris-Sud, 91405 Orsay, France

E-mail: raffy@lps.u-psud.fr

**Abstract.** We have epitaxially grown *c*-axis oriented $Sr_xLa_{1-x}CuO_2$ thin films by *rf* sputtering on $KTaO_3$ substrates with $x = 0.12$. The as-grown deposits are insulating and a series of superconducting films with various $T_c(\rho=0)$ up to 26 K have been obtained by in-situ oxygen reduction. Transport measurements in the *ab* plane of these samples have been undertaken. We report original results on the temperature dependence of the Hall effect and on the anisotropic magnetoresistance ($T \geq T_c$). We discuss the magnitude of upper critical fields and anisotropy, the Hall effect, which presents changes of sign indicative of the existence of two types of carriers, the normal state magnetoresistance, negative in parallel magnetic field, a possible signature of spin scattering. These properties are compared to those of hole-doped cuprates, such as *BiSr(La)CuO* with comparable $T_c$.

## 1. Introduction

$Sr_xLa_{1-x}CuO_2$ has the simplest structure among all the cuprates and is called infinite layer compound (IL) because it is only composed of an alternate stack of square $CuO_2$ planes and $Sr(La)$ planes. Studying a compound with such a simple structure is very attractive with the perspective to determine what the fundamental properties of cuprates are. By partial substitution of $Sr^{2+}$ by $La^{3+}$ cations it is possible to obtain an electron-doped compound [1]. Up to now essentially a single *e*-doped cuprate family in the form of thin films and single crystals, $Ln_{2-x}Ce_xCuO_4$ ($Ln = Nd, Pr...$), has been studied while a number of hole-doped cuprate families have been investigated in the last two decades. Comparing *e*-doped and *h*-doped cuprates is also an important issue to understand the superconductivity (SC) of cuprates.

The IL has been studied so far only in the form of ceramic samples, obtained by high pressure (2 GPa) synthesis [2], and measurements on oriented samples are lacking. No single crystal has been prepared yet as far as we know. However, by using epitaxy on an adequate substrate, it is possible to grow thin films [3], although it remains difficult to obtain SC samples. We have recently succeeded in preparing stable SC thin films [5] and been able to undertake a study of their transport properties. We will present results on the effect of the magnetic field on the resistive transition and on the magnetoresistance in perpendicular and parallel field for $T \geq T_c$. The results of the measurements of the Hall constant $R_H(T)$ as a function of temperature for $T \geq T_c$ are also described. It is interesting to compare these properties to the ones observed in an *h*-doped cuprate *BiSrCuO* and an *e*-doped *LnCeCuO* with comparable $T_c$.

---

[1] To whom any correspondence should be addressed.

## 2. Experimental details

Thin films of $Sr_{1-x}La_xCuO_2$ ($x = 0.12$) were prepared by single target *rf* magnetron sputtering on heated $KTaO_3$ substrates, a technique that we used for depositing *BiSrCuO* thin films [4]. The conditions of preparation are given in more details in reference 5. Since the as-prepared films are insulating, an oxygen reduction step is necessary to obtain SC films.

The films were mainly characterized by X-ray diffraction studies. Single phase epitaxial films were obtained as shown by $\theta$–$2\theta$ X-ray spectra [5]. The thickness of the films was in the range 40-60 nm. The mosaicity was of the order of 0.1°, even better than that of *BiSrCuO* thin films [4]. The composition was checked by inductive coupled plasma-mass spectroscopy (ICP) giving $x = 0.12$.

To perform transport measurements, the films were patterned in the shape of a conductor of small section between two large pads and equipped with two pairs of longitudinal contact pads. The track was typically 0.35 mm wide and 0.8 mm long. We used e-beam or optical lithography and chemical etching to pattern the samples. The transport measurements were performed for $H \leq 5$ T and $1.8 < T < 350$ K. The SC transition was checked both resistively and magnetically in a Quantum Design MPMS.

## 3. Results and discussions

### 3.1. Superconducting transition under magnetic field.

Figure 1 shows the resistivity curves as a function of temperature for five films with $x = 0.12$ and increasing oxygen content, denoted by numbers 1 to 5 respectively. The critical temperatures are defined either by the temperature where the resistivity goes to zero, $T_c(\rho=0)$, or by the temperature where $d\rho/dT$ is maximum called $T_{cmid}$. They decrease with decreasing electron doping (increasing oxygen content), the doping state being characterized by the conductivity at 300 K, $\sigma_{300K}$ (inset of figure 1). This kind of behavior is common for all underdoped cuprates. For all the samples studied, $\rho(T)$ curves have a positive curvature above 40 K. This curvature is weaker for the samples with higher $T_c$. It is to be reminded that, in the case of underdoped *BiSrLaCuO* films, the signature of the pseudogap is a downward deviation, below some temperature $T^*$, from the high $T$ linear behavior of $\rho(T)$ [6]. No such indication has been seen in $\rho(T)$ curves of our IL films. For the less doped samples (films 3, 4 and 5) and for decreasing temperature, the resistivity starts to increase below certain temperature $T_{min}$ and it reaches a maximum at temperature $T_p$ just before the SC transition (figure 1). The upturn in $\rho(T)$, found in *e*-doped *PrCeCuO*, was related to spin scattering mechanism [7] and the one found in *h*-doped *BiSrLaCuO* to the occurrence of disorder effects [6]. In the case of IL more detailed study is needed to resolve this question.

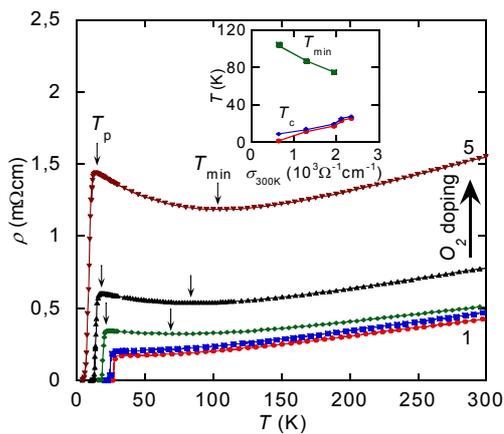

**Figure 1.** $\rho(T)$ for five films denoted by numbers 1 to 5. The long arrow indicates the direction in which the oxygen doping increases and small ones $T_{min}$ and $T_p$. The inset shows how $T_c(\rho=0)$, $T_{cmid}$ and $T_{min}$ vary with $\sigma_{300K}$.

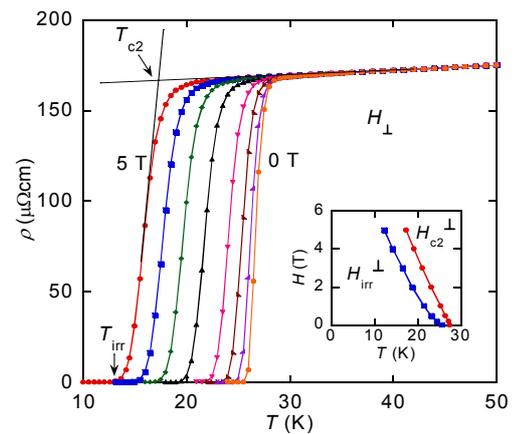

**Figure 2.** $\rho(T, H_\perp)$ curves of film 1, where $H_\perp = $ 5, 4, 3, 2, 1, 0.5, 0.2 and 0 T. Arrows indicate how the temperatures $T_{c2}$ and $T_{irr}$, where $H_{c2}^\perp$ and $H_{irr}^\perp$ set in, are determined. The inset shows $H_{c2}^\perp$ and $H_{irr}^\perp$ as a function of temperature.

Figure 2 displays a typical result of the effect of a perpendicular magnetic field, $H_\perp$, ($\mathbf{H} \perp ab$) on the resistive SC transition of film 1 with $T_c(\rho=0) = 26$ K. As illustrated in figure 2, the upper critical field $H_{c2}^\perp(T)$ is determined from the onset of the normal state at temperature $T_{c2}$ and the irreversibility field $H_{irr}^\perp(T)$ corresponds to the temperature $T_{irr}$ where the resistance goes to zero for a given value of the field. It appears that $H_\perp$ produces a shift of the transition to lower temperatures without significant broadening. As a result, $H_{irr}^\perp(T)$ is not very different from $H_{c2}^\perp(T)$ (inset of figure 2). In contrast, in the case of *h*-doped cuprates, such as *BiSrCuO*, a perpendicular magnetic field broadens the superconducting transition [8]. Using Werthamer-Helfand-Hohenberg theory [9] the critical field calculated at zero temperature is equal to $H_{c2}^\perp(T=0) = (10 \pm 2)$ T, giving a coherence length $\xi_{ab} = (6 \pm 1)$ nm, while a linear fit to the data, extrapolated to zero $T$, gives $H_{c2}^\perp(T=0) = (13.5 \pm 0.5)$ T and $\xi_{ab} = (4.9 \pm 0.2)$ nm. The anisotropy evaluated from a comparison of $\rho(H)$ curves at 26 K in perpendicular $H_\perp$ and parallel $H_\parallel$ ($\mathbf{H} \parallel ab$) magnetic field is at least equal to 14 (we had no facility to tune the magnetic field direction exactly parallel to *ab* plane). It means that the coherence length $\xi_c$ along the *c*-axis is larger than the distance $s = c = 0.34$ nm between the $CuO_2$ planes, which are well coupled. These findings are in good agreement with the ones found for polycrystalline samples of IL with $x = 0.1$ [10]. In *NdCeCuO* with $T_c = 24.4$ K, $H_{c2}^\perp(T=0) \sim 7$ T, $\xi_c \sim 0.3$ nm [11] and $s = c / 2 \cong 0.6$ nm and coupling between $CuO_2$ planes is weaker than for IL. In *BiSrCuO*, $s = c / 2 \cong 1.2$ nm, $\xi_c < 1$ nm and $H_{c2}^\perp(T=0) > 20$ T for films with $T_c = 15$ K [8].

*3.2. Hall effect in the normal state*

The linearity of Hall effect with the magnetic field ($H \leq 5$ T) was checked in the whole range of temperatures. The Hall effect is strongly dependent on temperature as shown by two typical $R_H(T)$ curves in figure 3. While less doped films (3, 4, 5) exhibit a single change of sign of Hall constant at $T_{high}^+$ (the lower $T_{high}^+$, the weaker the *e*-doping state), the films with higher $T_c$ exhibit two changes of sign, one at temperature $T_{high}^+$ and another at $T_{low}^+$ (inset of figure 3). A change of sign at low temperature was also reported in the other *e*-doped cuprate family. It indicates the presence of two types of carriers with different mobilities [11, 12]. In contrast no such change of sign has been seen in *BiSrCuO* thin films, where $\cot\theta_H = \rho / R_H H = A + BT^p$ (with p ≤ 2) is well defined [13]. The small magnitude of Hall constant may imply that the carrier concentration is significantly higher for IL than for *LnCeCuO* or *BiSrCuO* (~ $10^{21}$ cm$^{-3}$), due to a higher concentration of $CuO_2$ planes.

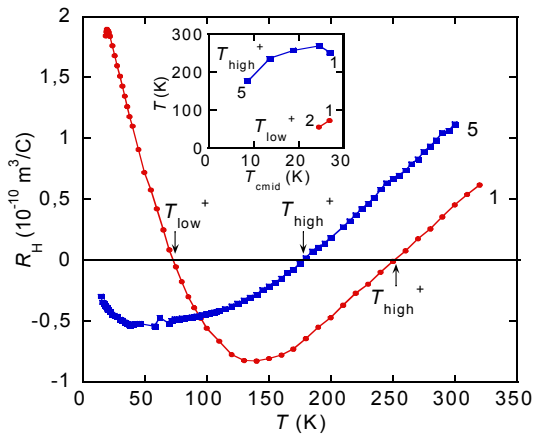

**Figure 3.** Temperature dependence of the Hall constant for films 1 and 5. The arrows indicate $T_{high}^+$ and $T_{low}^+$ where $R_H(T) = 0$. The inset shows $T_{high}^+$ and $T_{low}^+$ *vs.* $T_{cmid}$ of all the films studied.

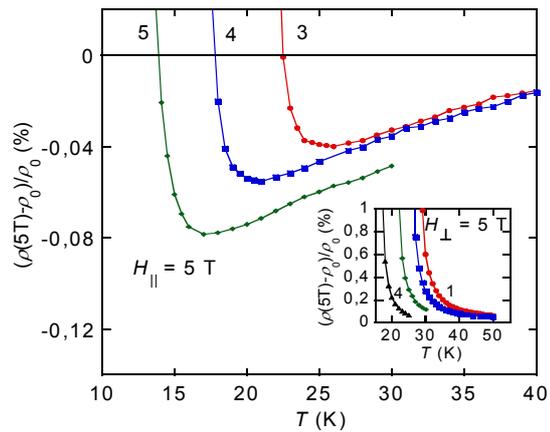

**Figure 4.** MR normalized by $\rho_0 = \rho(T, H=0)$ *vs.* temperature in $H_\parallel = 5$ T for films 3, 4 and 5. The inset shows normalized MR *vs.* temperature in $H_\perp = 5$ T for films 1, 2, 3 and 4.

*3.3. Magnetoresistance in the normal state*
Figure 4 shows the temperature dependence of magnetoresistance (MR) above $T_c$ in $H_\parallel$ = 5 T, and $H_\perp$ = 5 T in inset. In the range $T < 1.5T_c$, positive MR was obtained which is mainly attributed to the destruction of SC fluctuations by the field. In $H_\perp$, MR was positive for all temperatures. A quadratic $H_\perp$ dependence of MR at fixed temperature is observed in the normal state of several films, above $2T_c$ (not shown). Interestingly, in parallel field $H_\parallel$, a negative MR (nMR) is obtained for $T > T_p$, where $T_p$ is the temperature of the maximum in $\rho(T)$ just before the SC transition (figure 1). The magnitude of nMR is possibly higher than the one found because of eventual imperfect parallelism between field direction and *ab* plane. Also, it is to be noted that nMR has been observed in *LnCeCuO* thin films in the underdoped region of the phase diagram [14], where it was attributed to static AF and in *BiSrCuO* for very underdoped states [15], where it can be explained by e-e interactions and possibly by a spin scattering effect.

**4. Conclusion**
Single phase, *c*-axis oriented, epitaxial IL films with different oxygen doping were successfully grown on $KTaO_3$ substrates. $T_c$ and $\sigma_{300K}$ increase when oxygen doping decreases. At lower doping, an upturn in resistivity occurs and can possibly be related to AF or disorder. Perpendicular magnetic field shifts the transition towards lower temperatures with no significant broadening, indicating strong coupling between $CuO_2$ planes (anisotropy ~ 10). $H_{c2}^\perp(T=0)$ for the film with the highest $T_c$ (26 K) is larger than 10 T. Changes of sign of the Hall constant indicate the presence of two types of carriers, while the nMR in parallel field could lend support to the coexistence of AF and SC phases.

**Acknowledgments**
We wish to acknowledge Javier Briatico, CNRS/THALES, for optical and Raphael Weil, LPS, for e-beam lithography, Clarisse Mariet, CEA, Saclay for ICP analysis. One of us (V.J.) wishes to acknowledge the E.C. for a Marie Curie fellowship, contract number MEST-CT-2004-514307.